\documentclass[fleqn,usenatbib]{mnras}

\usepackage{newtxtext,newtxmath}

\usepackage[T1]{fontenc}
\usepackage{ae,aecompl}

\usepackage{graphicx}	
\usepackage{amsmath}	
\usepackage{amssymb}	
\usepackage{subcaption}
\usepackage{bm}
\usepackage[utf8]{inputenc}

\newcommand{\pix}{_{_{\rm PIX}}}

\title[Characterising lognormal fBm density fields with a CNN]{Characterising lognormal fractional-Brownian-motion density fields with a Convolutional Neural Network}

\author[M. L. Bates et al.]{
M. L. Bates,$^{1}$\thanks{E-mail: matthew.bates@astro.cf.ac.uk}
A. P. Whitworth$^{1}$
and O. D. Lomax,$^{1,2}$
\\
$^{1}$School of Physics and Astronomy, Cardiff University, Cardiff CF24 3AA, UK\\
$^{2}$ESTEC, Keplerlaan 1, 2201 AZ Noordwijk, Netherlands
}

\date{Accepted XXX. Received YYY; in original form ZZZ}

\pubyear{2019}

\begin{document}
\label{firstpage}
\pagerange{\pageref{firstpage}--\pageref{lastpage}}
\maketitle

\begin{abstract}
In attempting to quantify statistically the density structure of the interstellar medium, astronomers have considered a variety of fractal models. Here we argue that, to properly characterise a fractal model, one needs to define precisely the algorithm used to generate the density field, and to specify -- at least -- three parameters: one parameter constrains the spatial structure of the field; one parameter constrains the density contrast between structures on different scales; and one parameter constrains the dynamic range of spatial scales over which self-similarity is expected (either due to physical considerations, or due to the limitations of the observational or numerical technique generating the input data). A realistic fractal field must also be noisy and non-periodic. We illustrate this with the exponentiated fractional Brownian motion (xfBm) algorithm, which is popular because it delivers an approximately lognormal density field, and for which the three parameters are, respectively, the power spectrum exponent, $\beta$, the exponentiating factor, ${\cal S}$, and the dynamic range, ${\cal R}$. We then explore and compare two approaches that might be used to estimate these parameters: Machine Learning and the established $\Delta$-Variance procedure. We show that for $2\leq\beta \leq 4$ and $0\leq{\cal S}\leq 3$, a suitably trained Convolutional Neural Network is able to estimate  objectively both $\beta$ (with root-mean-square error $\epsilon_{_\beta}\sim 0.12$) and ${\cal S}$ (with $\epsilon_{_{\cal S}}\sim 0.29$). $\;\Delta$-variance is also able to estimate $\beta$, albeit with a somewhat larger error ($\epsilon_{_\beta}\sim 0.17$) and with some human intervention, but is not able to estimate ${\cal S}$.
\end{abstract}




\section{Introduction}\label{sec:Intro}

The interstellar medium is chaotic, due to the non-linear nature of the processes involved in its evolution (supersonic non-ideal magneto-hydrodynamics, self-gravity, radiation transport, non-LTE chemistry and heat transfer, etc.). Consequently the overall structure of the interstellar medium must be described using statistical metrics. Since in the interstellar medium there exist structures spanning a large dynamic range of spatial scales, and since there is evidence for self-similarity across parts of this dynamic range, there have been several attempts to characterise the interstellar medium, and in particular star forming clouds, with fractal or multi-fractal parameters \citep[e.g.][]{BeechM1987, BazellDesert1988, Falgaronetal1991, HeteLepi1993, stutzki_fractal_1998, BenschFetal2001, ChapScal2001, SanchezNetal2005, Ossenkopetal2008a, Kauffmanetal2010, Schneideetal2013, EliaDetal2014, Rathbornetal2015, EliaDetal2018}. Such characterisations can, in principle, allow one (a) to constrain the three dimensional structures and dynamics that underlie the observed two-dimensional projections; (b) to evaluate whether two observed regions might be statistically similar, even if their detailed structures are quite different; and (c) to compare the results of numerical simulations with observations, and with one another.

A variety of fractal metrics has been deployed. Of these, the conceptually simplest are the perimeter-area dimension, ${\cal D}_{_{\rm PA}}$ \citep[e.g.][]{BeechM1987, BazellDesert1988, Falgaronetal1991, HeteLepi1993, SanchezNetal2005, Federratetal2009, Rathbornetal2015}, and the box-counting dimension, ${\cal D}_{_{\rm BC}}$ \citep[e.g.][]{SanchezNetal2005, Federratetal2009, EliaDetal2018}; ${\cal D}_{_{\rm PA}}$ is usually preferred to ${\cal D}_{_{\rm BC}}$, because it tends to give less noisy results. A second group of metrics derive from structure, or structure-like, functions \citep[e.g.][]{SanchezNetal2005, Federratetal2009, KritsukAetal2013}; we include in this group the $\Delta$-Variance metric \citep{stutzki_fractal_1998, Ossenkopetal2008b, Federratetal2009}, which is the metric used here to compare with our CNN procedure. A third group of metrics involves evaluation of the  mass-length scaling relation \citep[e.g.][]{ChapScal2001, SanchezNetal2005, Federratetal2009, Kauffmanetal2010, KritsukAetal2013, BeattieJetal2019a, BeattieJetal2019b}. A fourth group involves estimating the size and/or mass spectra \citep[e.g.][]{ElmeFalg1996}, or the density spectrum \citep[e.g.][]{Federratetal2009, Konstandetal2016}.

There are two commonly used procedures for calibrating these metrics, and they are quite distinct. One procedure is based on idealised models of fractals generated using recursive algorithms. \citet{HeteLepi1993} describe three possible recursive fractal models, but do not identify a preferred model. \citet{SanchezNetal2005} use a model proposed by \citet{SoneirPeeble1978}, but have to adjust this model for high fractal dimensions. \citet{stutzki_fractal_1998}, \citet{Elmegree2002} and \citet{ShadElme2011} use models based on fractal Brownian motion (fBm) and exponentiated fractal Brownian motion (xfBm), and these are the models that we use here. 

Recursive fractal models have the problem that they are numberless, in the sense that there is no obvious limit to the possibility of inventing plausible new ones. One important distinction between different recursive fractal models is that some of them deliver nested fractals (i.e. fractals in which the smaller denser structures tend to be embedded within the larger more diffuse structures) and some do not. We have chosen the xfBm model because it delivers a lognormal density distribution: lognormal column-density in 2D, as here, and lognormal volume-density in 3D. Lognormal volume-density and column-density fields are commonly observed or inferred in (relatively) low-density gas  \citep[e.g][]{Schneideetal2012, Schneideetal2013, Kainulaietal2014}, and are usually attributed to compressible turbulence \citep[e.g.][]{VazquezS1994, Federratetal2010}. However xfBm does not yield a nested fractal; structures on different scales are positioned randomly with respect to one another. In a future paper we will explore whether nested fractal models provide a better model of the interstellar medium

Recursive fractal models for the interstellar medium require the specification of at least three parameters. One parameter reflects the relative frequency and spatial distribution of structures on different scales; here, this parameter is the power-law exponent, $\beta$, but it might equally be the fractal dimension, ${\cal D}$, or the Hurst parameter, ${\cal H}$. A second parameter reflects the way in which the density varies with physical scale; here this parameter is the exponentiating factor, ${\cal S}$ (as defined in Section \ref{SSEC:Exp}) -- but in other algorithms it is the Larson scaling exponent (i.e. $d\ln[\rho]/d\ln[L]$, where $\rho$ is density, and $L$ is a generic length-scale). A third parameter reflects the dynamic range of spatial scales, ${\cal R}$, over which the model is applied. This dynamic range might be determined by physical considerations, as for example in the theory of turbulence, which spans an inertial range from the large scales on which turbulent energy is injected to the small scales on which it is dissipated \citep[e.g.][]{FrischUr1995, FederrathC2013}. Alternatively the dynamic range of spatial scales might simply be determined by the limitations of the observations (between the field of view and the resolution of the telescope), or the limitations of the numerical technique (between the size of the computational domain and the smallest cell or particle); this is the case here.

The other procedure for calibrating fractal metrics is quite different, and is based on simulations of turbulent -- and usually non self-gravitating -- interstellar gas \citep[e.g.][]{Federratetal2009, KritsukAetal2013, Konstandetal2016, BeattieJetal2019a, BeattieJetal2019b}. The turbulence is maintained in isotropic statistical equilibrium with a random forcing term. Turbulent simulations are normally multi-fractal, firstly because the simulations have a limited dynamic range of spatial scales (and hence the turbulence has a limited inertial range), and secondly because the balance of solenoidal and compressive modes tends to depend on scale (with a shift from solenoidal to compressive modes as the turbulent energy cascades to smaller scales). The validity of this procedure depends on the fidelity of the simulations, on whether all the appropriate physics has been included, and on whether the real interstellar medium subscribes to isotropic statistical equilibrium.

Turbulent fractal simulations are usually characterised by just two parameters, the mean Mach Number, ${\cal M}$ of the turbulent velocity field, and the resolved dynamic range, ${\cal R}$. However, such simulations are also influenced by the way in which turbulent energy is continuously injected (e.g. the mix of solenoidal and compressive modes), the thermal and chemical behaviour of the gas, and the importance of self-gravity. Indeed \citet{deVegaetal1996} argue that fractal structure could be the natural product of self-gravity, rather than turbulence.

In all cases (both recursive fractal models, and turbulent fractal simulations), different realisations of the same model (with the same model parameters) are obtained by using different random number seeds.

The plan of this paper is as follows. In Section 2 we describe how 2D xfBm fields are constructed. In Section 3 we apply $\Delta$-variance to the analysis of such fields. In Section 4 we train a Convolutional Neural Network to analyse the same fields. In Section 5 we compare the two approaches and summarise our conclusions.

\begin{figure}
\includegraphics[width=\columnwidth]{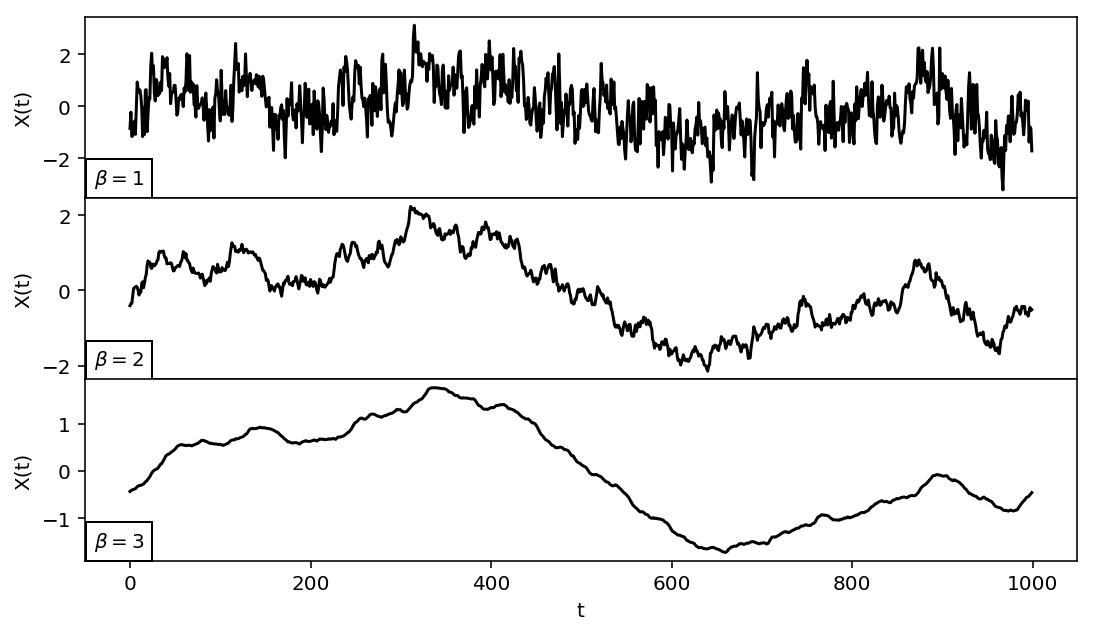}
\caption{One-dimensional (${\cal E}=1$) pure fBm curves for $\beta=1.0,\,2.0\;{\rm and}\;3.0$.}
\label{fig:1DSignal}
\end{figure}

\section{Constructing \texorpdfstring{\MakeLowercase{xf}B\MakeLowercase{m}}{xfBm} fields}\label{SEC:fBm}

xfBm fields are based on pure fBm fields, which we generate using the spectral synthesis method described by \cite{peitgen_science_1988}. The same methods have been used by \citet{stutzki_fractal_1998}, to create artificial molecular clouds, and by \citet{lomax_modelling_2018}, to create artificial star clusters.

Pure fBm fields ({\it un}-exponentiated fBm fields) are a generalised form of Brownian Motion and are characterised by a power-law spectrum with exponent $\beta = {\cal E} + 2{\cal H}$, where ${\cal E}$ is the Euclidean dimension and ${\cal H}$ is the Hurst parameter. Figure \ref{fig:1DSignal} demonstrates how a one-dimensional (${\cal E}=1$) pure fBm field, $X(t)$, depends on $\beta$. Each field has been realised with the same random seed, in order to preserve the general shape, but larger $\beta$ means more power on larger scales, and hence a smoother field. Irrespective of the value of $\beta$, the mean of the field over a sufficiently long $t$-interval, $\mu_{_X}$, is normally much smaller in magnitude than its standard deviation, $\sigma_{_X}$. The case $\beta=2.0$ corresponds to a 1D random walk, which is also sometimes described as classical Brownian motion.

For the rest of this paper we will work in two-dimensions (${\cal E}=2$), and hence we will be considering surface density fields. However, the procedures we discuss can easily be adjusted to treat other Euclidean dimensions. The methodology we use to create xfBm fields comprises five distinct stages; Stages 2 through 4 can be implemented in any order, but Stage 1 is always implemented first, and Stage 5 is always implemented last.

\subsection{Stage 1, generating a pure fBm field}\label{SSEC:Pure}

\noindent A pure fBm field, $f_{_\beta}(\bm r)$, is constructed by first generating a power spectrum $\hat{f}_{_\beta}(\bm k)$. Here, $\bm{r}\!\equiv\!(r_{_1},r_{_2})$ is a two-dimensional grid of integers with values of $1\leq r_{_i}\leq N\pix$, along each Cartesian axis, and $\bm{k}\!\equiv\!(k_{_1},k_{_2})$ is a two-dimensional grid of wave-vectors with integer values of $-N\pix/2\leq k_{_j}\leq N\pix/2$ along each Cartesian axis. For each $\bm k$, the contribution to the power spectrum is given by
\begin{eqnarray}
\hat{f}_{_\beta}(\bm k)&=&A_{_\beta}(\bm k)\,\left\{\vphantom{x^y_z}\cos(\varphi\bm{k})+i\sin(\varphi\bm{k})\,\right\};\\\label{eq:amplitudes}
A_{_\beta}(\bm k)&=&\left\{\begin{array}{ll}
0\,,&{\rm if}\;\bm{k}\,=\,\bm{0}\,;\\
{\cal K}^{-1/2}||\bm{k}||^{-\beta/2},&{\rm if}\;\bm{k}\,\neq\,\bm{0}\,;\\
\end{array}\right.\\
{\cal K}&=&\sum_{\bm k} \left\{||\bm k||^{-\beta}\right\}\,;\\\label{eq:phases}
\varphi(\bm k)&=&\chi(\bm k) - \chi(-\bm k);
\end{eqnarray}
$A_{_\beta}(\bm k)$ and $\varphi(\bm k)$ are, respectively, the amplitude and phase of the contribution. The normalisation factor ${\cal K}$ scales the total power of the field to unity. $\chi(\bm k)$ is a random variate sampled from a uniform distribution on the interval $0\leq\chi(\bm k)\leq 2\pi$. The pure fBm field, $f_{_\beta}(\bm r)$, is obtained by taking the inverse Fourier Transform of $\hat f_{_\beta}(\bm k)$.

\begin{figure*}
\includegraphics[width=1.0\linewidth]{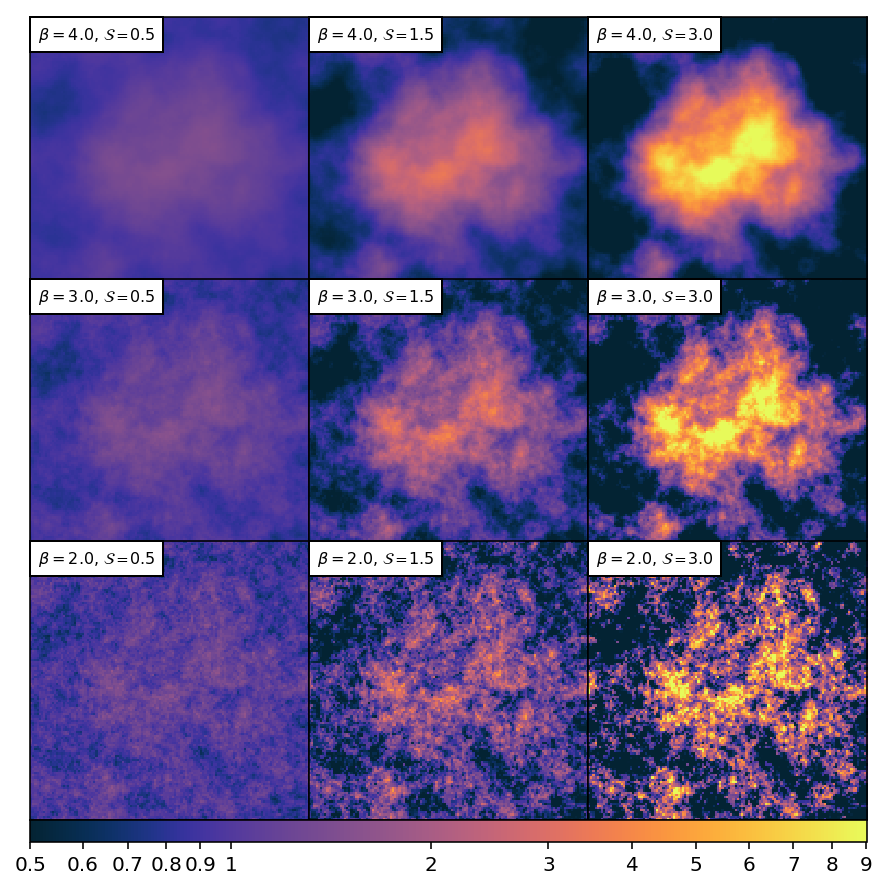}
\caption{2D fBm fields generated using the same random seed, but with different $\beta$ and ${\cal S}$, as shown in the top left corner of each panel. The fields are periodic and consist of $128\times128$ pixels. The logarithmic colour scale is the same in each plot, and shows the relative surface density, in arbitrary units.}
\label{fig:2DSignal}
\end{figure*}

\subsection{Stage 2, the exponentiated fBm field}\label{SSEC:Exp}

\noindent A pure fBm field has a roughly Gaussian distribution with a mean of around zero, $\langle f_{_\beta}(\bm r)\rangle \approx 0$. This means that roughly half of the field has negative values. In order to make the field everywhere positive, so that it can be used to model surface-density, we follow \citet{Elmegree2002} and exponentiate $f_{_\beta}(\bm r)$ to obtain an xfBm field,
\begin{eqnarray}\label{eq:exponentiated}
g_{_{{\cal HS}}}(\bm r)&=&\exp\left\{\frac{{\cal S}\; f_{_\beta}(\bm r)}{\left<f^2_{_\beta}(\bm r)\right>^{1/2}}\right\}\,,
\end{eqnarray}
using a scaling parameter, ${\cal S}$. $\;{\cal S}\!=\!0$ gives uniform density, and, as ${\cal S}$ is increased, the range of densities widens, and hence the structures become more sharply defined. This process transforms the roughly Gaussian field into one with a roughly lognormal distribution.

\subsection{Stage 3, the non-periodic xfBm field}\label{SSEC:NonPer}

\noindent The xfBm field, $g_{_{{\cal HS}}}(\bm r)$, is periodic, but observed fields are not. Therefore, we initially generate a pure fBm field, $g'_{_{{\cal HS}}}(\bm r')$, with $1\leq r'_{_i}\leq 4N\pix$, along each Cartesian axis. Then we cut out an $N\pix\times N\pix$ section, located so that its geometric centre coincides with its centre of mass. 

\subsection{Stage 4, the noisy xfBm field}\label{SSEC:Noisy}

 Since real observations are noisy, we add white noise to $g_{_{{\cal HS}}}(\bm r)$. The white noise field is scaled to be a fraction $\eta=0.05{\cal B}$ of the standard deviation, $\sigma_g$, of $g_{_{{\cal HS}}}(\bm r )$, where ${\cal B}$ is a linear random deviate on the interval $[0,1]$; hence the noise always lies between $0\%$ and $5\%$ of $\sigma_g$. This is only intended to be illustrative, but it is worth noting that higher noise levels will compromise $\Delta$-Variance more than the CNN.

\subsection{Stage 5, Adjusting rogue pixels}\label{SSEC:Rogue}

Finally, in order to filter out rogue pixels (which in a real map might represent cosmic ray strikes, for example), we compute the mean, $\mu_{_g}$, and standard deviation, $\sigma_{_{\!g}}$, for all pixels. Any pixels with $g>\mu_{_g}+2.5\sigma_{_{\!}g}$ are replaced with $\mu_{_g}+2.5\sigma_{_{\!g}}$, and similarly any pixels with $g<\mu_{_g}-2.5\sigma_{_{\!g}}$ are replaced with $\mu_{_g}-2.5\sigma_{_{\!g}}$. This cull of the most extreme pixels helps to stabilise the training and implementation of the CNN. Moreover, in observed clouds, the lognormal part of the column-density PDF is seldom well defined outside $\pm 2.5\sigma_{\!g}$, due to incompleteness on the low side, and a power-law tail \citep[usually attributed to self-gravity,][]{Girichidetal2014} on the high side \citep[e.g.][]{Schneideetal2012}.

\subsection{xfBm fields}\label{SSEC:Real}

Figure \ref{fig:2DSignal} shows how the appearance of an xfBm field, generated by the procedure outlined in the preceding sections, depends on $\beta$ and ${\cal S}$. These fields have all been generated from the same random seed in order that they all have the same large-scale pattern.

For the three fields on the top row, $\beta\!=\!4.0$ (equivalently ${\cal H}\!=\!1.0)$, the power is strongly concentrated in long-wavelength modes, and there is little small-scale structure; the same contours could be overlaid on all three images, albeit at different column-densities, and these contours would tend to be very smooth. For the three fields on the bottom row, $\beta\!=\!2.0$ (equivalently ${\cal H}\!=\!0.0)$, there is a lot of power at short wavelengths, and hence lots of small-scale structure; again, the same contours could be overlaid on all three images on the bottom row, albeit at different column-densities, and these contours would tend to be very twisted.

 For the three fields in the lefthand column, ${\cal S}\!=\!0.5$, the range of densities is the same and rather small. The only difference is that at the top ($\beta\!=\!4.0$) the density peaks and troughs are quite extended, and at the bottom ($\beta\!=\!2.0$) they are more compact. For the three fields in the righthand column, ${\cal S}\!=\!3.0$, the range of densities is also the same, but now it is rather big. Once again the density peaks and troughs at the top ($\beta\!=\!4.0$) are quite extended, and those at the bottom ($\beta\!=\!2.0$) are more compact. The range $0.5\leq{\cal S}\leq 3.0$ is chosen because this covers the range of variances in the column-density PDFs of observed clouds \citep[e.g.][]{Schneideetal2012, Schneideetal2013, Kainulaietal2014}.

\citet{stutzki_fractal_1998} show that the corresponding Box-Counting fractal dimension should be ${\cal D}_{_{\rm BC}}\simeq(3{\cal E}+2-\beta)/2$, and the corresponding Perimeter-Area fractal dimension should be ${\cal D}_{_{\rm PA}}\simeq(3{\cal E} -\beta)/2={\cal D}_{_{\rm BC}}-1$, where ${\cal E}$ is the Euclidean dimension, and we have used `$\simeq$' because our xfBm fields are not pure. Substituting ${\cal E}\!=\!2$, we obtain ${\cal D}_{_{\rm BC}}\simeq (8-\beta)/2$ and ${\cal D}_{_{\rm AP}}\simeq (6-\beta)/2$.

In Figure \ref{fig:2DSignal}, and in the rest of the paper, we use $N\pix=128$, so the dynamic range of spatial scales is ${\cal R}\la\, 64$. In the next two Sections we explore two techniques for characterising xfBm fields constructed in this way: $\Delta$-Variance (Section \ref{SEC:DeltaVar}) and Convolutional Neural Networks (Section \ref{SEC:CNN}).

\section{$\Delta$-variance} \label{SEC:DeltaVar}

The $\Delta$-variance, $\sigma^2_\Delta(L)$, of a 2D field, $g(x,y)$ is the variance after the field has been convolved with a circular filter function, $\odot_{_L}$, characterised by length-scale $L$:
\begin{eqnarray}
\sigma_\Delta^2(L)&=&\frac{1}{2\pi} \left\langle\left(g*\odot_{_L}\right)^2\right\rangle_{x,y}.
\end{eqnarray}
$\sigma_\Delta^2(L)$ must be evaluated for many different values of $L$, spanning the full dynamic range of spatial scales being modelled. The power-law exponent, $\beta$, is then given by
\begin{eqnarray}\label{EQN:Hurst}
\beta&=&{\cal E}\,+\,\frac{d\ln\left(\sigma_\Delta^2\right)}{d\ln(L)}
\end{eqnarray}
In computing this gradient, care must be taken to discount end effects, i.e. where $L$ is either close to the scale of the whole field, or close to the resolution limit; this issue is discussed further in Section \ref{SSEC:periodic} below.

In the original formulation \citep{stutzki_fractal_1998}, the French Hat filter function has been used, but \citet{Ossenkopetal2008b} show that better results are obtained with the Mexican Hat filter function,
\begin{eqnarray}
\odot_{_L}(r)\!\!&\!\!=\!\!&\!\!\odot_{_{{\rm CORE.}L}}(r)-\odot_{_{{\rm ANN.}L}}(r)\,,\\
\odot_{_{{\rm CORE.}L}}(r)\!\!&\!\!=\!\!&\!\!\frac{4}{\pi L^2}\,\exp\left(\frac{-\,4\,r^2}{L^2}\right), \\
\odot_{_{{\rm ANN.}L}}(r)\!\!&\!\!=\!\!&\!\!\frac{4}{\pi(\nu^2-1)L^2}\left\{\exp\left(\frac{-\,4\,r^2}{\nu^2\,L^2}\right)-\exp\left(\frac{-\,4\,r^2}{L^2}\right)\right\}\!,\hspace{0.75cm}
\end{eqnarray}
and this is the filter which we use here.

\subsection{The power-law exponent, $\beta$, for periodic fields}\label{SSEC:periodic}

\begin{figure}
\includegraphics[width=\columnwidth]{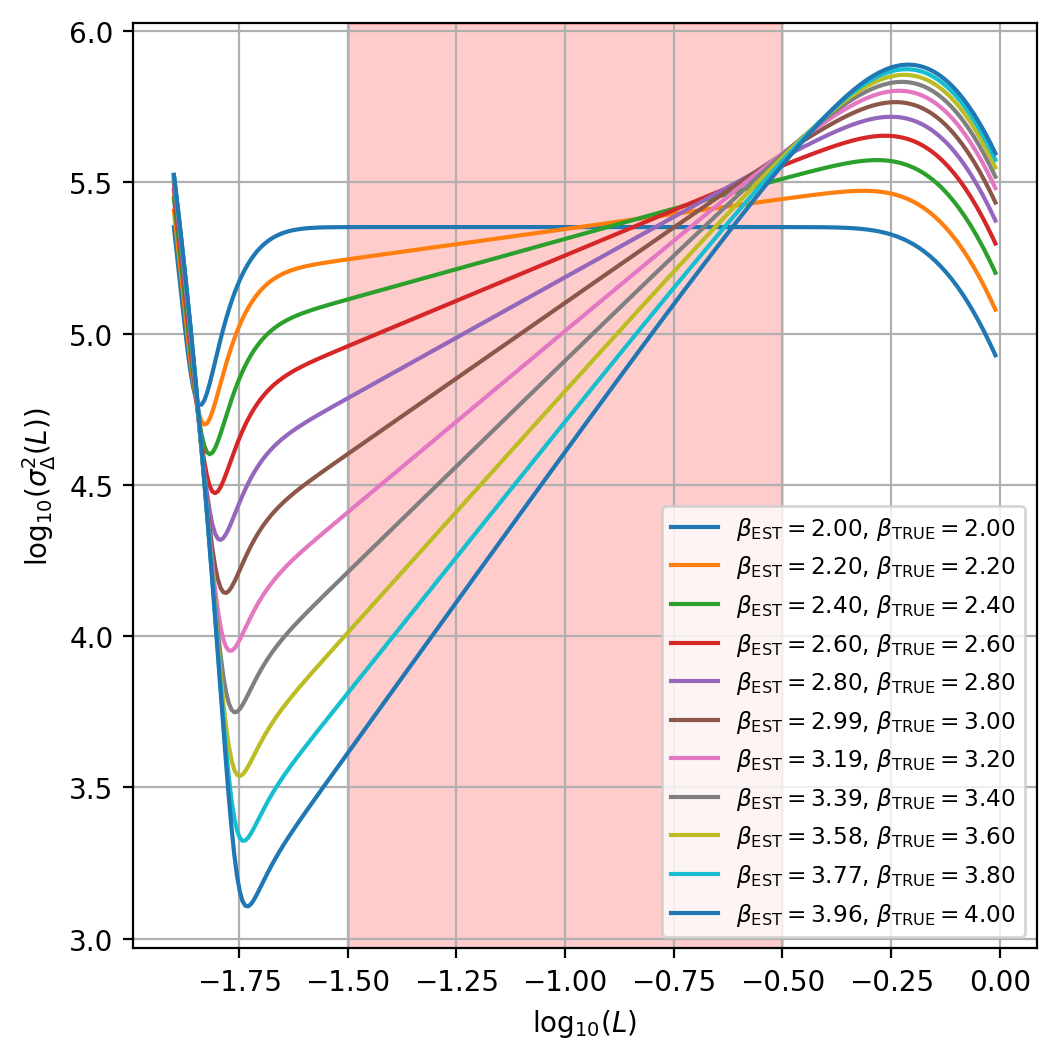}
\caption{$\Delta$-variance curves for pure (i.e. periodic, un-exponentiated and noiseless) fBm fields with power-law exponent $\beta_{_{\rm TRUE}} = 2.0,\, 2.2,\, 2.4,\, 2.6,\, 2.8,\, 3.0,\, 3.2,\, 3.4,\, 3.6,\, 3.8\;{\rm and}\; 4.0$ (hence ${\cal H}_{_{\rm TRUE}} = 0.0,\, 0.1,\, 0.2,\, 0.3,\, 0.4,\, 0.5,\, 0.6,\, 0.7,\, 0.8,\, 0.9\;{\rm and}\; 1.0$). The pink shading shows the range used to estimate the slope, and hence (using Eqn. \ref{EQN:Hurst}) to obtain $\beta_{_{\rm EST}}$. Values of $\beta_{_{\rm TRUE}}$ and $\beta_{_{\rm EST}}$ are tabulated in the corner of the frame.}
\label{fig:DVCurves}
\end{figure}

As noted by \citet{Ossenkopetal2008b}, for periodic fields (but only for periodic fields), the $\Delta$-variance can be computed more quickly by integrating the product of the power spectrum of $g$ (denoted ${\cal P}_{_{\!g}}\!(\bm k)$) and the power spectrum of the filter function (denoted $\tilde\odot_{_L}$) over $\bm k$-space,
\begin{eqnarray}
\sigma_\Delta^2(L)&=&\frac{1}{2\pi}\int{\cal P}_{_{\!g}}\!(\bm k)\,\left|\tilde\odot_{_L}\right|^2\,d^2\bm{k}\,.
\end{eqnarray}
Figure \ref{fig:DVCurves} shows the $\Delta$-variance curves obtained in this way for eleven pure (i.e. periodic, un-exponentiated and noise-less) fBm fields with $\beta_{_{\rm TRUE}} = 2.0,\, 2.2,\, 2.4,\, 2.6,\, 2.8,\, 3.0,\, 3.2,\, 3.4,\, 3.6,\, 3.8\,{\rm and}\, 4.0$. If we limit consideration to the range $-1.50\leq \log_{_{10}}(L)\leq-0.50$ (shaded pink on Fig. \ref{fig:DVCurves}), the slope is in all cases well defined, and can be used to estimate $\beta$ from Eqn. (\ref{EQN:Hurst}). The values estimated in this way, $\beta_{_{\rm EST}}$, are tabulated in the corner of Fig. \ref{fig:DVCurves}, and agree well with the input values, $\beta_{_{\rm TRUE}}$.

\subsection{The power-law exponent, $\beta$, for non-periodic fields}\label{SSEC:nonperiodic}

\citet{Ossenkopetal2008b} also note that for non-periodic fields, a more convoluted procedure is required. First, the map is zero-padded to twice the linear size. Next, the convolution is performed, using the original filter size, but only on the pixels which constitute the original map, in order to prevent the filter from wrapping around the edges of the map. This involves four convolution integrals,
\begin{eqnarray}
G_{_{{\rm CORE.}L}}(\bm r)&=&g_{_{\rm PAD}}(\bm r')*\odot_{_{{\rm CORE.}L}}(\bm r)\,,\\
G_{_{{\rm ANN.}L}}(\bm r)&=&g_{_{\rm PAD}}(\bm r')*\odot_{_{{\rm ANN.}L}}(\bm r)\,,\\
W_{_{{\rm CORE.}L}}(\bm r)&=&w(\bm r')*\odot_{_{{\rm CORE.}L}}(\bm r)\,,\\
W_{_{{\rm ANN.}L}}(\bm r)&=&w(\bm r')*\odot_{_{{\rm ANN.}L}}(\bm r)\,,
\end{eqnarray}
where $g_{_{\rm PAD}}(\bm r')$ is the zero-padded map, and $w(r')$ is a normalisation map which takes values of $1$ within the region of the original map, and $0$ in the zero-padded region. The fully-convolved map is then computed using
\begin{eqnarray}
F_{_L}(\bm r) = \frac{G_{_{{\rm CORE.}L}}(\bm r)}{W_{_{{\rm CORE.}L}}(\bm r)}-\frac{G_{_{{\rm ANN.}L}}(\bm r)}{W_{_{{\rm ANN.}L}}(\bm r)}\,,
\end{eqnarray}
and the $\Delta$-variance is given by
\begin{eqnarray}
\sigma^2_\Delta(L)&=&\frac{\sum\left\{(F_{_L}(\bm r)-\langle F_{_L}(\bm r)\rangle)^2\,W_{_{{\rm TOT.}L}}(\bm r)\right\}}{\sum \left\{W_{_{{\rm TOT.}L}}(r)\right\}}\,.\hspace{0.5cm}
\end{eqnarray}
Here, $W_{_{{\rm TOT.}L}}(\bm r) = W_{_{{\rm CORE.}L}}(\bm r)\,W_{_{{\rm ANN.}L}}(\bm r)$ acts as a map of weights, which, when applied to the variance calculation, gives less significance to the pixels that are most heavily distorted by edge effects due to the zero-padding.

\begin{figure*}
  \begin{subfigure}[b]{0.5\linewidth}
    \centering
    \includegraphics[width=0.75\linewidth]{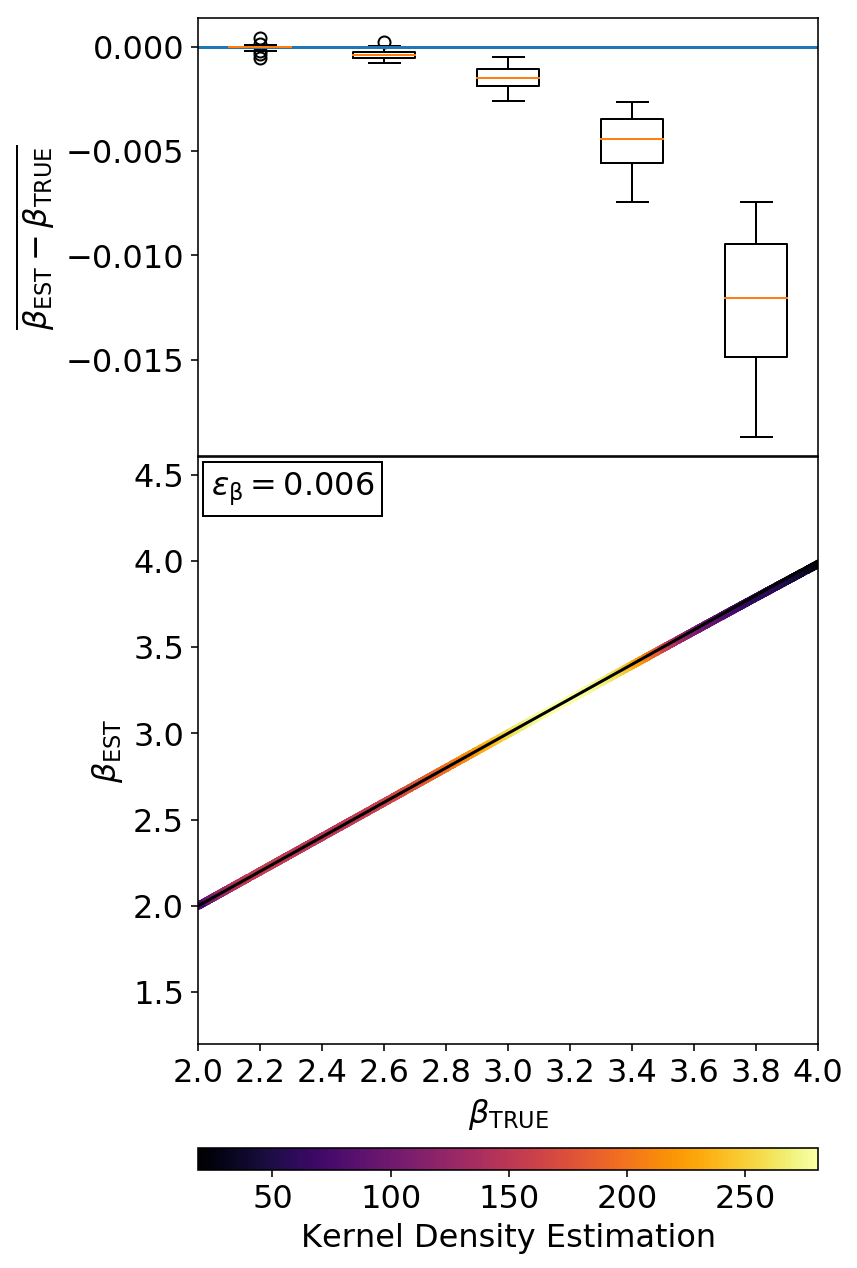} 
    \caption{Pure}
\label{fig:DVPerfa} 
    \vspace{4ex}
  \end{subfigure}
  \begin{subfigure}[b]{0.5\linewidth}
    \centering
    \includegraphics[width=0.675\linewidth]{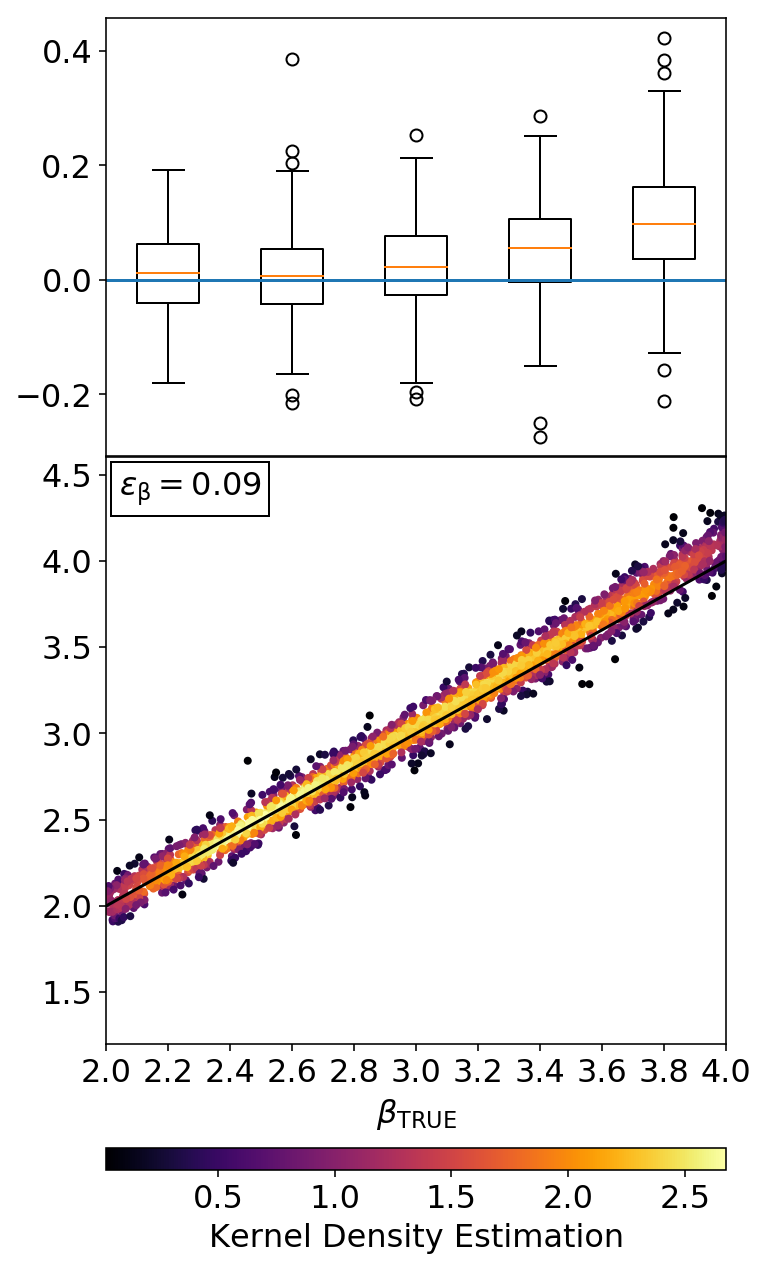} 
    \caption{Non-Periodic} 
\label{fig:DVPerfb} 
    \vspace{4ex}
  \end{subfigure} 
  \begin{subfigure}[b]{0.5\linewidth}
    \centering
    \includegraphics[width=0.72\linewidth]{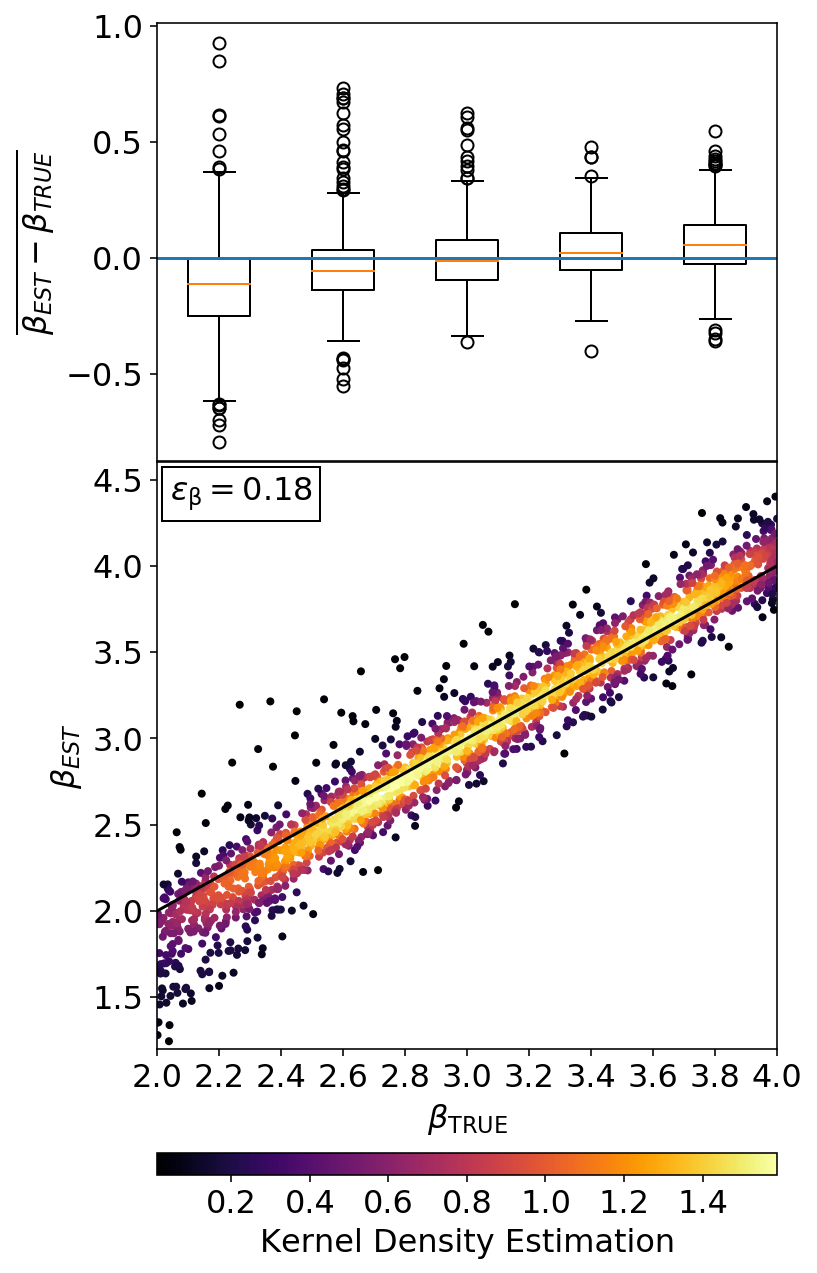} 
    \caption{Exponentiated} 
\label{fig:DVPerfc} 
  \end{subfigure}
  \begin{subfigure}[b]{0.5\linewidth}
    \centering
    \includegraphics[width=0.68\linewidth]{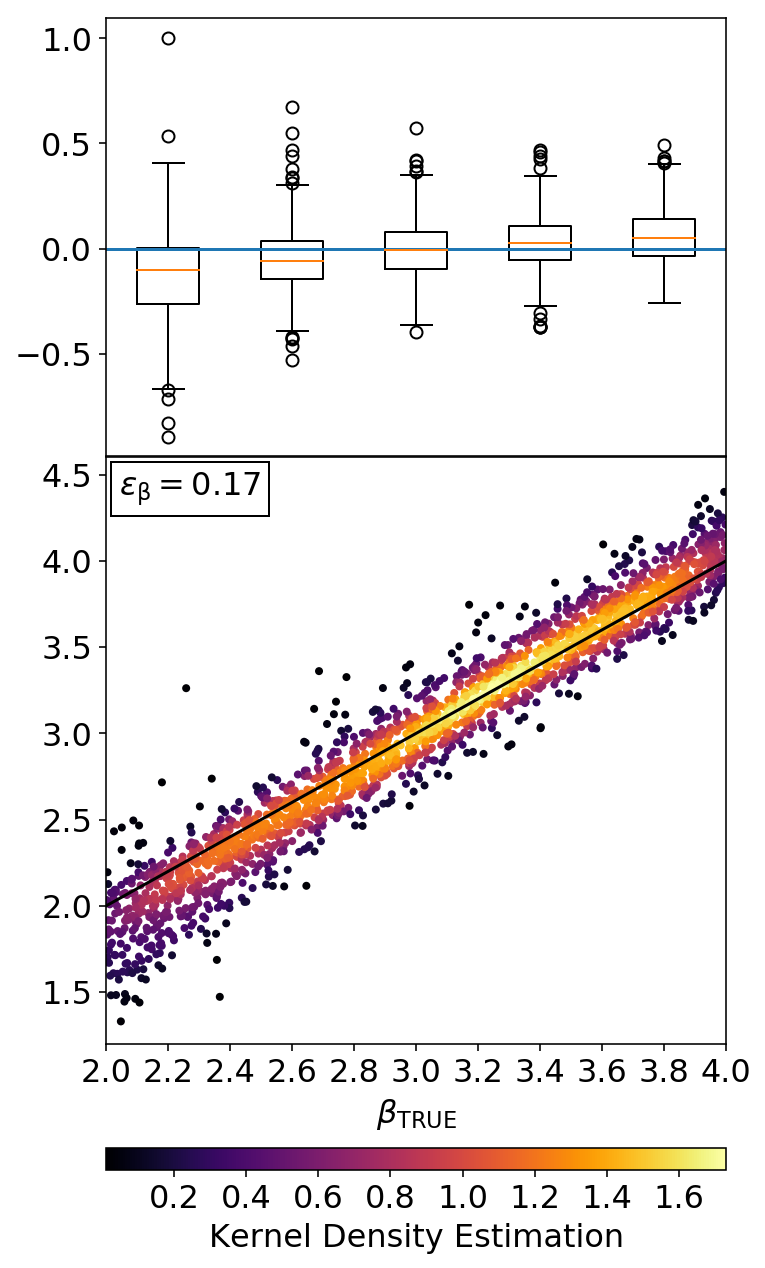} 
    \caption{Noisy} 
\label{fig:DVPerfd} 
  \end{subfigure} 
   \caption{Comparison of the input values of the power-law exponent, $\beta_{_{\rm TRUE}}$, with the values estimated using $\Delta$-variance, $\beta_{_{\rm EST}}$. The box-and-whisker plots on the top row show the distribution of $(\beta_{_{\rm EST}}\!\!-\beta_{_{\rm TRUE}})$ in bins of width $\Delta\beta_{_{\rm TRUE}}\!=\!0.4$. In each bin the orange line marks the median, and the box spans from the lower quartile, $Q_1$, to the upper quartile, $Q_3$. If the interquartile range is $\Delta Q=Q_3-Q_1$, the upper whisker extends to the highest point less than $Q_3\!+\!1.5\Delta Q$, the lower whisker extends to the lowest point greater than $Q_1\!-\!1.5\Delta Q$, and all points outside this range are plotted individually as open circles. The blue line marks exact correspondence. The kernel density estimates on the bottom row show the correspondence between $\beta_{_{\rm EST}}$ and $\beta_{_{\rm TRUE}}$, and $\epsilon_{_\beta}$ is given, for each stage, in the top lefthand corner of the panel. Reading from left to right and top to bottom, the plots correspond to (a) {\it pure} fBm fields, (b) {\it non-periodic} fBm fields, (c) {\it exponentiated}, non-periodic fBm fields, and (d) {\it noisy}, exponentiated, non-periodic fBm fields.}
\label{fig:DVPerf} 
\end{figure*}

\subsection{Evaluating the performance of $\Delta$-variance}\label{SSEC:DVLimitations}

To test the above procedures, we use the methodology outlined in Section \ref{SEC:fBm} to construct 2000 different artificial xfBm fields, each measuring $128\times 128$ pixels, and each with a random value of $\beta_{_{\rm TRUE}}$ on the interval $[2.0\,,4.0]$, and a random value of ${\cal S}_{_{\rm TRUE}}$ on the interval $[0\,,3]$. At each stage in the construction, we apply $\Delta$-variance to estimate  $\beta_{_{\rm EST}}$, and compare the result with $\beta_{_{\rm TRUE}}$. The results are presented in Fig. \ref{fig:DVPerf}. Note that for this exercise we have reversed the order of Stages 2 and 3 (Sections \ref{SSEC:Exp} and \ref{SSEC:NonPer}).

For the pure fBm fields generated in Stage 1 (Section \ref{SSEC:Pure}), we are able to use the procedure for periodic fields outlined in Section \ref{SSEC:periodic}, and the same range ($-1.50\leq \log_{_{10}}(L)\leq-0.50$). The results are presented in Fig. \ref{fig:DVPerfa}. In this case there is almost exact correspondence between $\beta_{_{\rm EST}}$ and $\beta_{_{\rm TRUE}}$. The root-mean-square error is $\epsilon_{_\beta}\simeq 0.006\,$.

For the non-periodic fields generated in the subsequent stages (Sections \ref{SSEC:Exp} to \ref{SSEC:Noisy}), we have to use the more convoluted procedure for treating non-periodic fields, as outlined in Section \ref{SSEC:nonperiodic}, and consequently the estimates of the power-law exponent deteriorate. Fig. \ref{fig:DVPerfb} shows the results obtained with non-periodic fBm fields; in this case $\epsilon_{_\beta}\simeq 0.09$, and there is a tendency to overestimate $\beta_{_{\rm EST}}$ for high values of $\beta_{_{\rm TRUE}}$. Fig. \ref{fig:DVPerfc} shows the results obtained for exponentiated non-periodic fields; in this case $\epsilon_{_\beta}\simeq 0.18$, and there is still a tendency to overestimate $\beta_{_{\rm EST}}$ for high values of $\beta_{_{\rm TRUE}}$, but also a tendency to underestimate $\beta_{_{\rm EST}}$ for low values of $\beta_{_{\rm TRUE}}$. The addition of noise does not change the error significantly, i.e. it is $\epsilon_{_\beta}\simeq 0.17\,$.

\begin{figure}
\includegraphics[width=\columnwidth]{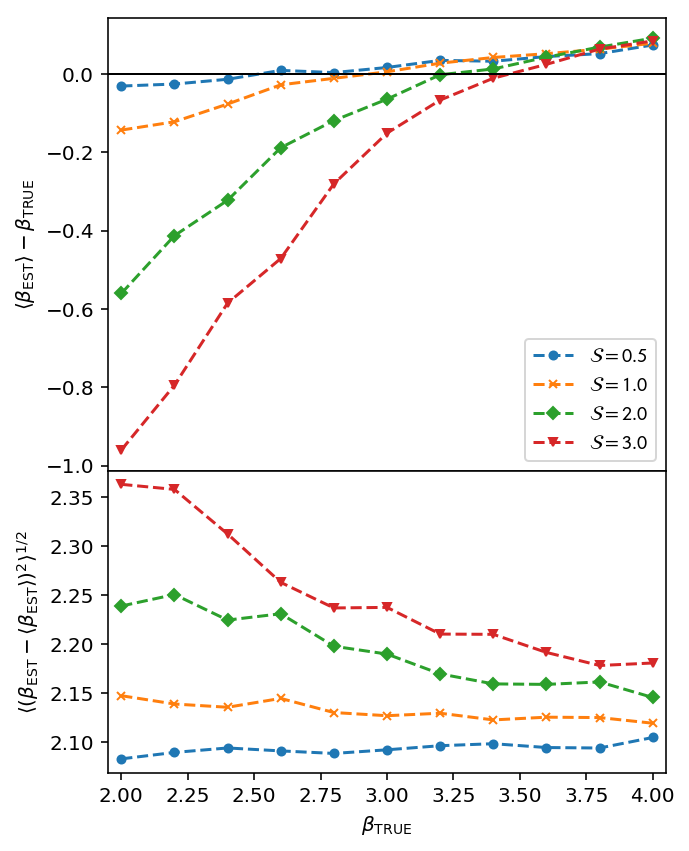}
\caption{Values of $\,\langle\beta_{_{\rm EST}}\rangle\!-\!\beta_{_{\rm TRUE}}\,$ (top panel) and $\langle(\beta_{_{\rm EST}}-\langle\beta_{_{\rm EST}}\rangle)^2\rangle^{1/2}$ (bottom panel) for discrete values of $\beta_{_{\rm TRUE}} = 2.0,\, 2.2,\, 2.4,\, 2.6,\, 2.8,\, 3.0,\, 3.2,\, 3.4,\, 3.6,\, 3.8\;{\rm and}\; 4.0$, and discrete values of ${\cal S}_{_{\rm TRUE}} =0.5$ (filled blue circles), $1.0$ (orange crosses), $2.0$ (filled green diamonds), and $3.0$ (filled red triangles). For each combination of $\beta_{_{\rm TRUE}}$ and ${\cal S}_{_{\rm TRUE}}$, $\beta_{_{\rm EST}}$ has been estimated using $\Delta$-variance. The mean and standard deviation of $\beta_{_{\rm EST}}$ are based on 400 different artificial xfBm fields (i.e. non-periodic, exponentiated, noisy fields).}
\label{fig:SigmaDV}
\end{figure}

In order to explore the interplay between the parameters $\beta$ and ${\cal S}$ and how this is reflected in the values of $\beta_{_{\rm EST}}$ obtained using $\Delta$-variance, we consider discrete values of $\beta_{_{\rm TRUE}} = 2.0,\, 2.2,\, 2.4,\, 2.6,\, 2.8,\, 3.0,\, 3.2,\, 3.4,\, 3.6,\, 3.8\;{\rm and}\; 4.0$ and ${\cal S}_{_{\rm TRUE}} = 0.5,\, 1.0,\, 2.0 \;{\rm and}\; 3.0$. Then, for each combination of $\beta_{_{\rm TRUE}}$ and ${\cal S}_{_{\rm TRUE}}$, we generate 400 different artificial xfBm fields (i.e. non-periodic, exponentiated and noisy); estimate their individual $\beta_{_{\rm EST}}$ using $\Delta$-variance; and hence determine the mean, $\mu_{_\beta}=\langle\beta_{_{\rm EST}}\rangle$, and standard deviation, $\sigma_{_\beta}=\langle(\beta_{_{\rm EST}}-\langle\beta_{_{\rm EST}}\rangle)^2\rangle^{1/2}$. Fig. \ref{fig:SigmaDV} displays the results. In general, as ${\cal S}_{_{\rm TRUE}}$ increases, the mean, $\mu_{_\beta}$, falls increasingly far below $\beta_{_{\rm TRUE}}$, and the standard deviation, $\sigma_{_\beta}$, increases. These trends are particularly strong for low values of $\beta_{_{\rm TRUE}}$. Inspection of Fig. \ref{fig:2DSignal} suggests that these trends arise because increasing ${\cal S}$ and reducing $\beta$ both have the effect of amplifying the visibility of small-scale structure in the field. $\Delta$-variance is unable to distinguish these two effects, as noted previously by \citet{lomax_modelling_2018}.

\section{Convolutional Neural Networks} \label{SEC:CNN}

The use of neural networks for classification and regression has expanded rapidly in recent years. A large variety of different types of network has emerged, most notably the Convolutional Neural Network (CNN), which is used extensively in problems involving image recognition. A notable example is handwritten digit recognition \citep{ciresan_flexible_2011, ciresan_multi-column_2012}. Several competitions have also served to push the boundaries of CNNs, for instance the annual ImageNet Large Scale Visual Recognition Challenge (ILSVRC), which in 2012 established the usefulness of Graphic Processing Units when combined with deep CNNs  \citep{krizhevsky_imagenet_2017}.

More recently machine learning techniques have started to be applied to problems in astronomy. Examples of the use of CNNs include galaxy classification \citep{khalifa_deep_2017}, gamma-ray astronomy \citep{dieleman_rotation-invariant_2015,postnikov_particle_2018}, supernova classification \citep{kimura_single-epoch_2017}, astronomical image reconstruction \citep{flamary_astronomical_2016}, denoising of images \citep{remez_deep_2017}, and star cluster analysis \citep{bialopetravicius_deriving_2019}.

\begin{table}
\centering
\caption{The architecture of the CNN. The initial $128\times128\times1$ input layer is the 2D field to be analysed, and the final $1\times1\times2$ output layer gives the estimated $\beta$ and ${\cal S}$. In between there are 5 convolutional layers, each followed by a max pooling function, and then 5 flattened, fully connected, dense layers. The Output Size column follows the format: \texttt{width$\times$height$\times$channels}. The total number of parameters is $11,545,090$.}
\label{tab:architecture}
\begin{tabular}{lcc}\hline
Layer & Output Size & Operation\\\hline
Input & $128\times128\times1$ & input layer\\
Conv.1 & $126\times126\times512$ & $3\times3$ kernel\\
MaxPool.1 & $63\times63\times512$ & $2\times2$ max pooling\\
Conv.2 & $61\times61\times512$ & $3\times3$ kernel\\
MaxPool.2 & $30\times30\times512$ & $2\times2$ max pooling\\
Conv.3 & $28\times28\times512$ & $3\times3$ kernel\\
MaxPool.3 & $14\times14\times512$ & $2\times2$ max pooling\\
Conv.4 & $12\times12\times512$ & $3\times3$ kernel\\
MaxPool.4 & $6\times6\times512$ & $2\times2$ max pooling\\
Conv.5 & $4\times4\times512$ & $3\times3$ kernel\\
MaxPool.5 & $2\times2\times512$ & $2\times2$ max pooling\\
Flatten & $1\times1\times2048$ & flattens into 1D layer\\
Dense.1 & $1\times1\times512$ & fully connected\\
Dense.2 & $1\times1\times512$ & fully connected\\
Dense.3 & $1\times1\times512$ & fully connected\\
Dense.4 & $1\times1\times512$ & fully connected\\
Dense.5 & $1\times1\times512$ & fully connected\\
Output & 1$\times1\times2$ & one channel each for $\beta$ and ${\cal S}$\\\hline
\end{tabular}
\end{table}

\subsection{Architecture of the CNN}

\begin{figure*}
  \begin{subfigure}[b]{0.5\linewidth}
    \centering
    \includegraphics[width=0.75\linewidth]{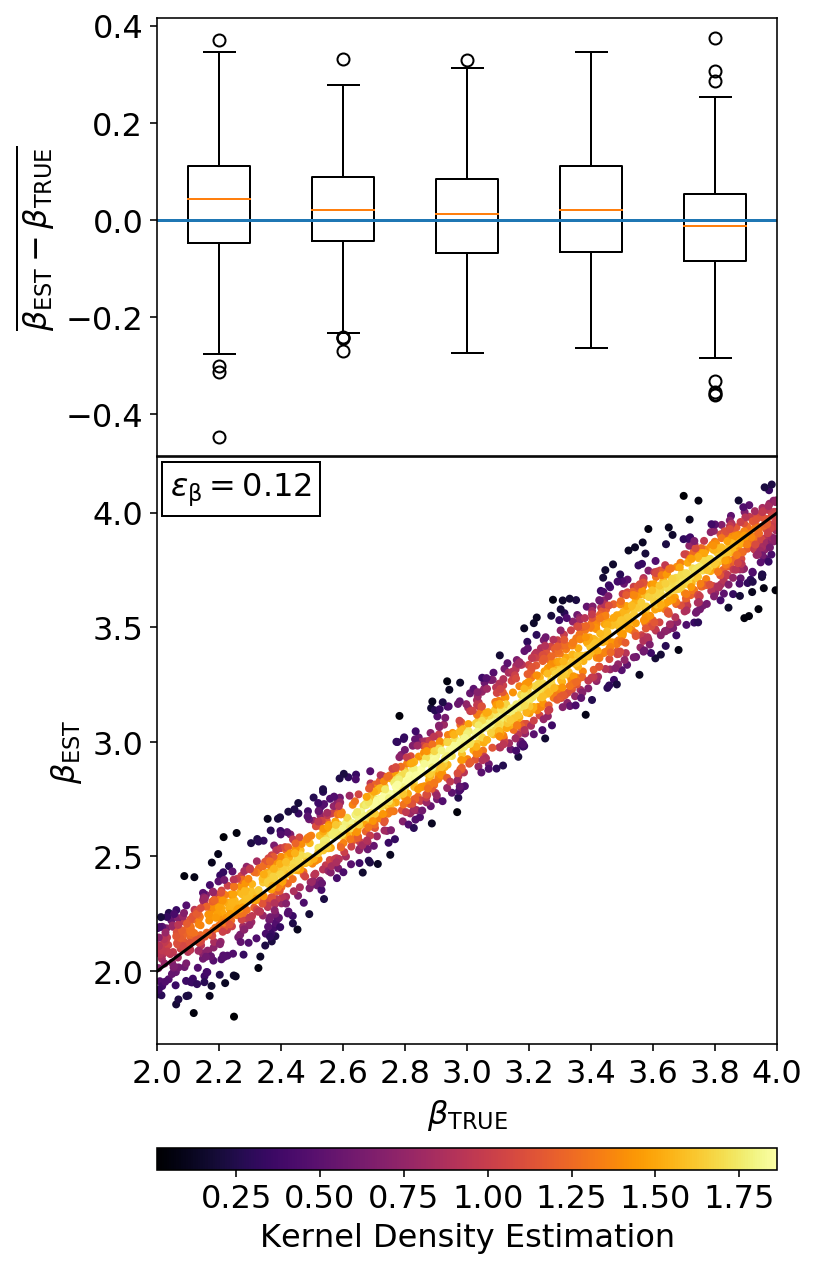} 
    \caption{$\beta$}
\label{fig:NNPerfH} 
    \vspace{4ex}
  \end{subfigure}
  \begin{subfigure}[b]{0.5\linewidth}
    \centering
    \includegraphics[width=0.75\linewidth]{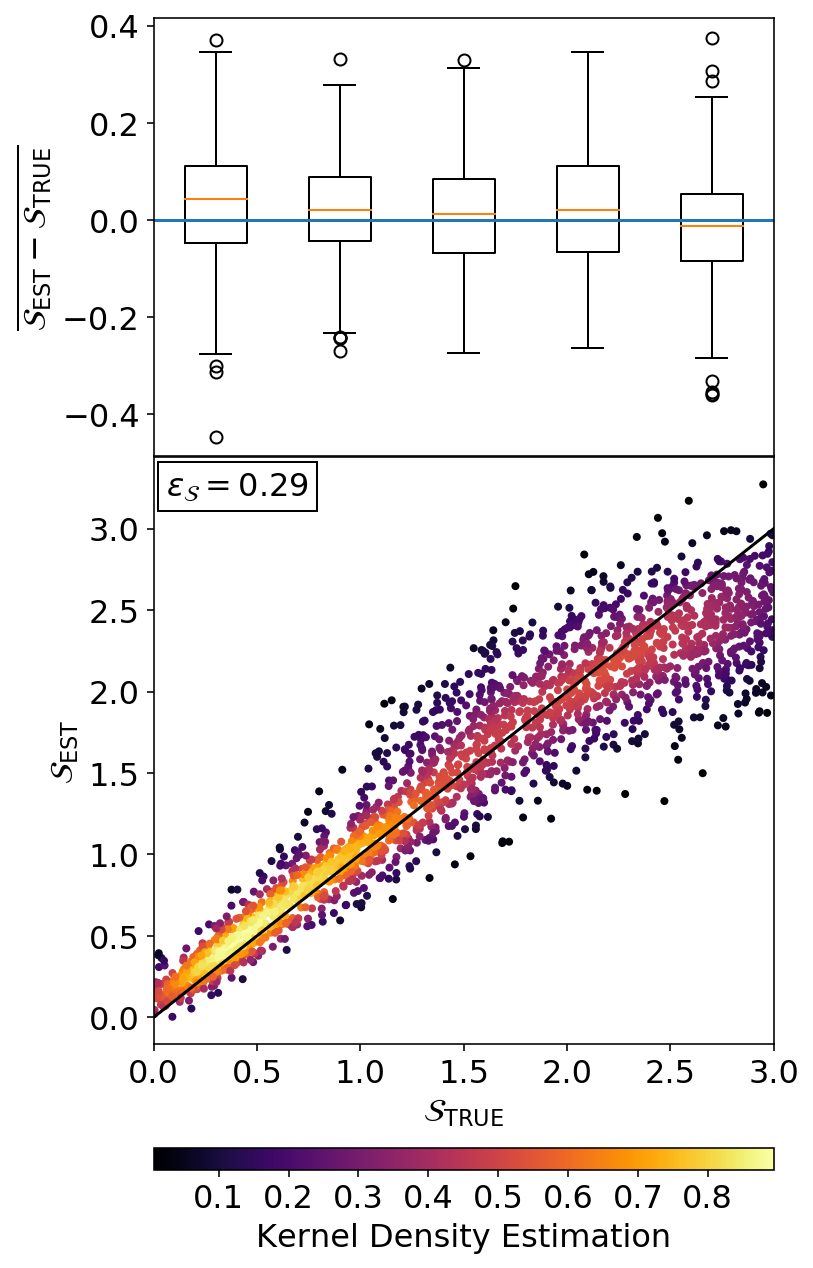} 
    \caption{${\cal S}$} 
\label{fig:NNPerfSigma} 
    \vspace{4ex}
  \end{subfigure} 
    \caption{{\it Left panels:} comparison of the input values of the power-law exponent, $\beta_{_{\rm TRUE}}$ with the values returned by the CNN, $\beta_{_{\rm EST}}$. {\it Right panels:} comparison of the input values of the scaling factor, ${\cal S}_{_{\rm TRUE}}$ with the values returned by the CNN, ${\cal S}_{_{\rm EST}}$. The box-and-whisker plots on the top row show the distribution of $(\beta_{_{\rm EST}}\!\!-\beta_{_{\rm TRUE}})$ in bins of width $\Delta\beta_{_{\rm TRUE}}\!=\!0.4$, and $({\cal S}_{_{\rm EST}}\!\!-{\cal S}_{_{\rm TRUE}})$ in bins of width $\Delta{\cal S}_{_{\rm TRUE}}\!=\!0.6$. In each bin the orange line marks the median, and the box spans from the lower quartile, $Q_1$, to the upper quartile, $Q_3$. If the interquartile range is $\Delta Q=Q_3-Q_1$, the upper whisker extends to the highest point less than $Q_3\!+\!1.5\Delta Q$, the lower whisker extends to the lowest point greater than $Q_1\!-\!1.5\Delta Q$, and all points outside this range are plotted as open circles. The blue line marks exact correspondence. The kernel density estimates on the bottom row show the correspondence between $\beta_{_{\rm EST}}$ and $\beta_{_{\rm TRUE}}$, and ${\cal S}_{_{\rm EST}}$ and ${\cal S}_{_{\rm TRUE}}$; the values of $\epsilon_{_\beta}$ and $\epsilon_{_{\cal S}}$ are given in the top lefthand corner of each panel.}
\label{fig:NNPerf} 
\end{figure*}

A CNN consists of a collection of artificial neurons, with each neuron taking a vector of inputs $\bm x$, and producing a scalar output, $y = f(c+\bm w \cdot \bm x)$. Here, $\bm w$ is a vector of weights, $c$ is a bias, and $f(\cdot)$ is an activation function; the activation function used here is the Rectified Linear Unit (ReLU), $f(x) = \mbox{\sc max}[0,x]$ \citep{nair_rectified_2010}.

Neurons are arranged in multiple groupings called layers, and each neuron in the layer takes all the outputs from the previous layer as its inputs. In general, a layer delivers a vector of outputs $\bm y$; and a sequence of layers forms a neural network. Table \ref{tab:architecture} shows the structure of the CNN developed here, using the \verb|Tensorflow| package. It consists of 5 convolutional layers (Conv.N), each followed by a max pooling layer (MaxPool.N). These are then flattened into a 1-dimensional layer which is then followed by 5 fully connected layers (Dense.N).

The weights and biases of the neurons comprise the parameters of the network, and are refined using multiple sets of input data ($\bm x_{_{\rm INPUT}}$) and their corresponding known statistical parameters ($\bm y_{_{\rm KNOWN}}$). Gradient descent is then used to minimise a loss function $\mathcal L$, which we set to the mean square error,
\begin{eqnarray}
{\cal L}&=&\langle (F(\bm x_{_{\rm INPUT}})-\bm y_{_{\rm KNOWN}})^2\rangle\,.
\end{eqnarray}
Here, $F(\bm x_{_{\rm INPUT}})$ is the estimate of $\bm y$ delivered by the CNN.

The convolutional layers of the CNN consist of two-dimensional grids of multiple, learnable convolutional filters. Each filter comprises a $3\times 3$ window, made up of $9$ parameters. The window is moved across the map in steps, producing an output at each step by computing the dot product between the filter and the local subsection of the map. The CNN used here has 5 convolutional layers, each using 512 different filters (so that it produces 512 different feature maps) and a step of 1 (so that it reduces the size of the layer by 2 in each dimension). The 9 parameters for each filter are refined by minimisation of the loss function.

Each convolutional layer is followed by a max pooling layer, using a $2\times2$ window and a step of 2. Max pooling outputs the maximum value of a $2\times2$ subsection of the layer. The step of 2 means the window moves 2 pixels before outputting the next maximum, thereby halving the image size.

The input to the CNN is a single channel, $128\times 128$ pixel xfBm field. The first convolutional layer (Conv.1) produces 512 different feature maps, and these are then carried through the network, until they are condensed into 512 single neurons at the Dense.1 layer, and finally into 2 singular neurons at the Output layer, i.e. the values of $\beta_{_{\rm EST}}$ and ${\cal S}_{_{\rm EST}}$.

\subsection{Training the CNN}\label{subsubsec:Training}

To train the CNN we generate $20,000$ artificial xfBm fields (using the procedures described in Section \ref{SEC:fBm}), each with a random value of $\beta$ on the interval $[2.0,4.0]$, a random value of ${\cal S}$ on the interval $[0,3]$, and $128\times 128$ pixels. The CNN's parameters start out with random values. The artificial xfBm fields are then input to the network in batches of 32, the input $\beta_{_{\rm TRUE}}$ and ${\cal S}_{_{\rm TRUE}}$ are compared with the values estimated by the network, $\beta_{_{\rm EST}}$ and ${\cal S}_{_{\rm EST}}$, and the parameters updated using the \verb|RMSprop| gradient-descent optimiser, so as to minimise the loss function, ${\cal L}$. For a comprehensive review of different optimisers and their applicability see \cite{ruder_overview_2016}. We train the CNN for 100 epochs with a random 70-30 train-test cross-validation split. For details of this cross-validation split, see Appendix \ref{sec:A1}.

\subsection{Evaluating the performance of the CNN}

We test the performance of the CNN using the same 2000 artificial fBm fields that were used in Section \ref{SSEC:DVLimitations} to test the performance of $\Delta$-variance. Fig. \ref{fig:NNPerfH} shows that the CNN tends to overestimate the power-law-exponent, $\beta$, but the error is small, $\epsilon_{_\beta}=0.12$. Fig. \ref{fig:NNPerfSigma} shows that the CNN also tends to overestimate the scaling factor, ${\cal S}$, except for large values (${\cal S} >2$), which it tends to underestimate; the error is $\epsilon_{_{\cal S}}=0.29$.

\section{Discussion and conclusions}\label{sec:Dis}

It appears that the CNN developed here is able to estimate the power-law exponent, $\beta$, of an xfBm field (i.e. an fBm field that has been exponentiated, and is non-periodic and noisy) more accurately (rms error $\epsilon_{_\beta} =0.12$) than $\Delta$-variance ($\epsilon_{_\beta} =0.18$). In addition, the CNN can also evaluate the scaling factor (${\cal S}$) with reasonable accuracy ($\epsilon _{_{\cal S}}=0.29$). 

Training and cross-validating a CNN takes about four hours on a GPU cluster, but applying the CNN to  a single, $128\times 128$ pixel xfBm field then takes $\la\, 0.1\,{\rm sec}$. In contrast, $\Delta$-variance requires no training, but applying it to a single, $128\times 128$ pixel xfBm field takes $\sim 2\,{\rm secs}$ (on the same computer architecture), because it entails the computation of several convolution integrals over the whole field. It may also require human intervention to identify the range over which the plot of $\log_{_{10}}(\sigma^2_{_\Delta}(L))$ against $\log_{_{10}}(L)$ is linear.

The CNN developed here can only be applied to $128\times 128$ pixel fields. Given a field with $N_{_{\rm PIX}}\neq 128$, we have three choices. (i) We can convert the field to $128\times 128$ pixels. (ii) If $N_{_{\rm PIX}}>128$, we can divide the field up into $128\times 128$ pixel subfields, analyse each subfield separately, and combine the results with appropriate weights. (iii) We can develop a new CNN. In contrast, $\Delta$-variance can be applied immediately to a field with any number of pixels.

The disadvantage of both approaches is that they return parameter values irrespective of whether the fields being analysed are actually well approximated by fractional Brownian motion. This is particularly true for the CNN, which is a black box with no demonstrable relation to underlying physical structures. $\Delta$-variance can at least provide some (necessary but not sufficient) evidence for an underlying fBm structure, if the plot of $\log_{_{10}}(\sigma^2_{_\Delta}(L))$ against $\log_{_{10}}(L)$ displays a linear portion (as demonstrated for the pure fBm fields analysed in Fig. \ref{fig:DVCurves}), but this may require human intervention. It might therefore be appropriate to combine the two approaches: use the CNN to estimate $\beta_{_{\rm CNN}}$ and ${\cal S}_{_{\rm CNN}}$, and then re-estimate $\beta_{_{\rm \Delta -VAR}}$ using $\Delta$-variance, and check whether it falls below $\beta_{_{\rm CNN}}$ in accordance with the results of Fig. \ref{fig:SigmaDV}.

\section*{Acknowledgements}
MLB gratefully acknowledges the receipt of a PhD studentship from the UK Science and Technology Facilities Council (STFC) through the Centre for Doctoral Training (CDT) in Data Intensive Science (ST/P006779/1). APW and ODL gratefully acknowledge the support of an STFC Consolidated Grant (ST/K00926/1). ODL also gratefully acknowledges the support of an ESA Fellowship. This work was performed using the computational facilities of the Advanced Research Computing at Cardiff (ARCCA) Division, Cardiff University. We thank the referee for their careful report on the original version of this paper, which we found very helpful.

\bibliographystyle{mnras}
\bibliography{references}

\appendix

\section{Optimisation}\label{sec:A1}

\begin{figure}
\includegraphics[width=0.95\columnwidth]{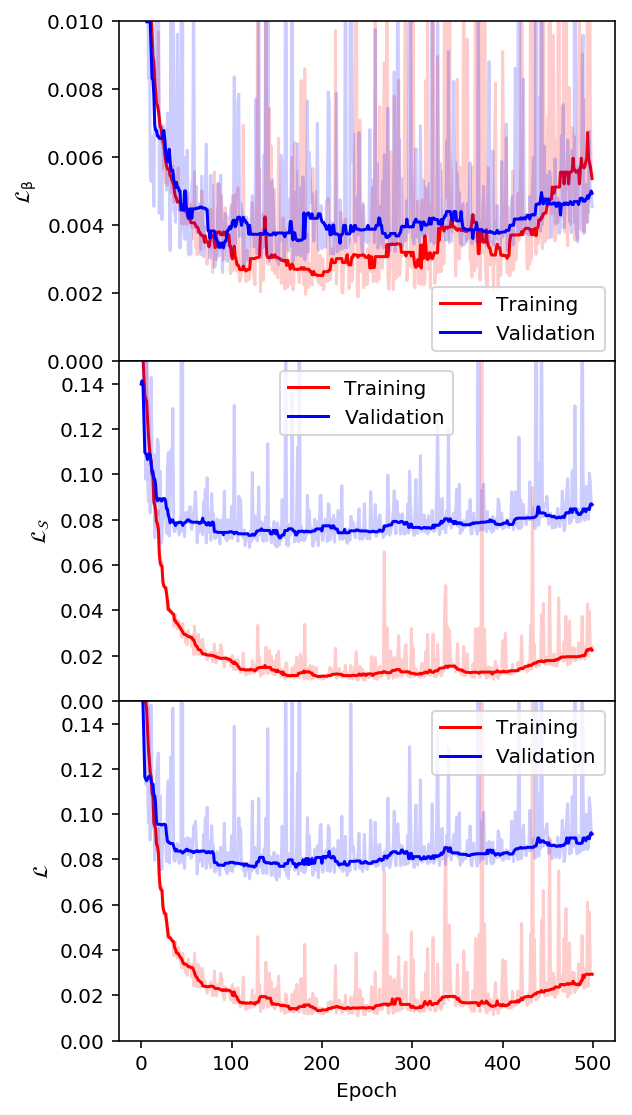}
\caption{The loss function for $\beta$ (${\cal L}_{_\beta}$, top panel); ${\cal S}$ (${\cal L}_{_{\cal S}}$, middle panel); and the total (${\cal L}={\cal L}_{_\beta}+{\cal L}_{_{\cal S}}$, bottom panel). The pale red (blue) curves show how the loss function for the training (testing) set evolves with epoch, and the dark red (blue) curves are smoothed versiosn obtained by taking the median of the 20 surrounding points.}
\label{fig:Loss}
\end{figure}

The CNN was initially trained for 500 full passes, or epochs, of the input dataset (the 20,000 artificial fBm fields).  At each epoch, a random 70\% of the artificial fields (i.e. 14,000 fields) were selected and used to train the network, by minimising the associated loss function, ${\cal L}_{_{\rm TRAIN.70\%}}$. The remaining 30\% (6,000 fields) were set aside and used to cross-validate the network, by computing its loss function, ${\cal L}_{_{\rm VALID.30\%}}$, separately. This cross-validation is designed to check that the network is not overfitting the dataset. If it is, ${\cal L}_{_{\rm VALID.30\%}}$ will tend to increase systematically with successive passes, while ${\cal L}_{_{\rm TRAIN.70\%}}$ will generally continue to decrease. We train for a large number of epochs (500) in order to determine the point at which the CNN starts to overfit. Fig. \ref{fig:Loss} shows the evolution of the {\sc train}.{\small 70\%} and {\sc valid}.{\small 30\%} loss functions. Separate plots are given for the contributions to the loss functions from $\beta$ and ${\cal S}$, and for their sum. We see by eye that ${\cal L}_{_{\rm VALID.30\%}}$ starts to increase at $\sim100$ epochs. Therefore we restrict the CNN to 100 epochs for the analyses described in Section \ref{SEC:CNN}.

After $\sim 400$ epochs, ${\cal L}_{_{\rm TRAIN.70\%}}$ also starts to increase. This suggests that the gradient-descent optimiser (here, \verb|RMSProp|) is taking too large a step and thus moving away from the loss-function minimum.

We tested the dependence on image size by repeating the analyses in Sections \ref{SEC:DeltaVar} and \ref{SEC:CNN} with $100\times100$ pixel xfBm images. Using a CNN there was no significant change in the accuracy, with $\epsilon_{_\beta} = 0.12$ and $\epsilon_{_{\cal S}} = 0.31$. Using $\Delta$-variance, the accuracy was somewhat worse, with $\epsilon_{_\beta} = 0.18$.

We also tested several distinct CNN architectures, and different numbers of layers and different numbers of nodes. The architecture described in the text (Table \ref{tab:architecture}) appears to deliver reasonable accuracy using modest computation time.

\bsp	
\label{lastpage}
\end{document}